\pdfoutput=1 
\documentclass[reprint, amssymb, amsmath, aip, jap, floatfix]{revtex4-1}

\usepackage[colorlinks=true,linkcolor=blue]{hyperref}
\usepackage[pdftex]{graphicx}
\usepackage{xspace}
\usepackage{bm}
\usepackage{float}

\newcommand{\dummySml}{\rule[0ex]{0pt}{3ex}}
\newcommand{\dummy}{\rule[-1ex]{0pt}{4ex}}
\newcommand{\dummyBtm}{\rule[-1.5ex]{0pt}{1.5ex}}

\newcommand{\AlInN}{\ensuremath{\rm{Al}_x\rm{In}_{1-x}\rm{N}}\xspace}
\newcommand{\GaInN}{\ensuremath{\rm{Ga}_x\rm{In}_{1-x}\rm{N}}\xspace}
\newcommand{\AlGaInN}{$\rm{Al_x}\rm{Ga}_{y}\rm{In}_{1-x-y}\rm{N}$\xspace}
\newcommand{\abinitio}{\emph{ab initio}\xspace}

\usepackage{type1cm,eso-pic}

\makeatletter

\AddToShipoutPicture*{\setlength{\@tempdimb}{0.0\paperwidth}\setlength{\@tempdimc}{1.0\paperheight}\setlength{\unitlength}{1pt}\put(\strip@pt\@tempdimb,\strip@pt\@tempdimc){\makebox( 430,-75){{\textcolor[gray]{0.50}{\fontsize{9pt}{9pt}
\selectfont{This is a preprint of the paper published in
\href{http://dx.doi.org/10.1063/1.3678002}
{J. Appl. Phys. 111, 033502 (2012).}
}}}
}}}
\makeatother

\begin{document}
\title[Composition dependence of elastic constants in wurtzite AlGaInN
alloys]{Composition dependence of elastic constants in wurtzite AlGaInN alloys}

\author{M. \L{}opuszy\'nski }

\affiliation{ 
Interdisciplinary Centre for Mathematical and Computational Modelling, 
University of Warsaw, Pawi{\'n}skiego 5A, 02-106 Warsaw, Poland
}
\email{m.lopuszynski@icm.edu.pl}

\author{J. A. Majewski}
\affiliation{
Institute of Theoretical Physics, Faculty of Physics,
University of Warsaw, Ho{\.z}a 69, 00-681 Warsaw, Poland
}


\begin{abstract}
In this paper, we analyze the dependence of elastic constants $c_{ij}$ on
composition for random wurtzite quaternary \AlGaInN alloy in the whole
concentration range. The study takes as its starting point the $c_{ij}$
parameters for zinc blende phase calculated earlier by the authors on the basis
of valence force field model. To obtain the wurtzite constants from cubic
material parameters the Martin transformation is used.  The deviations from
linear Vegard-like dependence of $c_{ij}$ on composition are analyzed and
accurate quadratic fits to calculated moduli are presented. The influence of
nonlinear internal strain term in the Martin transformation is also
investigated. Our general results for quaternary \AlGaInN alloys are compared
with the recent \abinitio calculations for ternaries \GaInN and \AlInN
(Gorczyca and \L{}epkowski 2011 {\it Phys. Rev.} B {\bf 83} 203201) and good
qualitative agreement is found.
\end{abstract}

\maketitle

\section{Introduction}
Nitrides are currently the most promising materials for blue, green, and UV
optoelectronics.  Their applications are diverse and include biosensors, medical
imaging, optical data storage, multimedia, etc.  One of the important methods of
controlling properties of these materials relies on alloying instead of
employing pure AlN, GaN, or InN.  Modern crystal growth techniques allow for
fabricating ternary (e.g., \GaInN) and even quaternary \AlGaInN mixtures of these
semiconductors.  By adjusting the composition, one can tune selected material
parameters such as bandgap, lattice constant or polarization to the desired
optimal value.  It is worth stressing that quaternary alloys, having two
compositional degrees of freedom, exhibit much greater tuning potential.  In
this material, it is possible to control not only the band-gap alone, but also
independently another property. Sample application of quaternary alloys'
flexibility involve controlling both band-gap and polarization charges, which
leads to so-called polarization-matched quantum wells showing very good
performance when applied in laser diodes. \cite{Schubert2008}  Therefore, the
question about the properties of quaternary alloys is vital.  Unfortunately, for
these materials not much is known about the exact composition dependence of many
properties. Among such quantities are elastic constants.  They are very useful
when modeling strained alloy layers in quantum heterostructures by means of
continuous or coarse grained models (e.g.,  $\bm{k} \cdot \bm{p}$ theory). For
sample applications for quaternary \AlGaInN, see, e.g., Refs.
[\onlinecite{Soh2005}] and [\onlinecite{Sakalauskas2011}]. So far, in such
modeling, linear dependence of elastic properties on alloy composition was
routinely assumed (Vegard-like law). However, there are serious indications that
deviations from this simple rule should be expected in the case of elastic
properties, \cite{Chen1995, Lopuszynski2007a, Lopuszynski2010, Lepkowski2011} 
similarly to, e.g., nonlinear dependence on composition predicted for
piezoelectric properties \cite{Ambacher2002}. Since subtle nonlinear effects in
elasticity and electrostriction proved to be significant when modeling nitride
devices, \cite{Lepkowski2005, Lepkowski2008, BahramiSamani2010}  also the
accurate composition dependence of $c_{ij}$ can be of interest, for correct and
accurate description of nanostructures. In the present study, we give a detailed
overview of influence of composition on $c_{ij}$ in random wurtzite \AlGaInN.
The work is based on our previous calculations for cubic phases.
\cite{Lopuszynski2010}  The paper is organized as follows. In
Sec.~\ref{sec:Methods}, we give an overview of the employed methodology and
compare the results obtained for binary materials AlN, GaN, and InN with
available experimental values.  In Sec.~\ref{sec:Results}, we present the
results obtained for composition dependence of $c_{ij}(x,y)$ in hexagonal
\AlGaInN alloys. We also include there a comparison of our results with recent
\abinitio calculations of elastic constants in ternary \GaInN and \AlInN.
Section~\ref{sec:Summary} concludes the paper.

\section{Martin Transformation and Elastic Constants for Wurtzite 
         Nitrides \label{sec:Methods}}
In our previous work, \cite{Lopuszynski2010} we calculated the composition
dependence of elastic constants in zinc blende \AlGaInN alloys using valence
force field (VFF) approach. \cite{Keating1966}  It is well known that zinc
blende and wurtzite structures are closely related. Many semiconducting
materials exhibit zinc blende--wurtzite polytypism. \cite{Yeh1992} Moreover,
many properties of the cubic and hexagonal phases resemble each other. In the
seventies, Martin derived a transformation between the three independent elastic
constants for cubic materials $c^{\text{zb}}_{11}$, $c^{\text{zb}}_{12}$,
$c^{\text{zb}}_{44}$ and five independent constants of wurtzite
$c^{\text{w}}_{11}$, $c^{\text{w}}_{12}$, $c^{\text{w}}_{13}$,
$c^{\text{w}}_{33}$, $c^{\text{w}}_{44}$. \cite{Martin1972,Martin1979} This
transformation can be done in two steps. In the first stage simple linear
relation is employed
\begin{equation}
\left(
\begin{array}{r} 
  c^{\text{w0}}_{11}  \\ 
  c^{\text{w0}}_{33}  \\ 
  c^{\text{w0}}_{12}  \\
  c^{\text{w0}}_{13}  \\
  c^{\text{w0}}_{44}  
\end{array} \right)
=\frac{1}{6}
\left(
\begin{array}{rrr} 
   3  &  3  &  6  \\ 
   2  &  4  &  8  \\ 
   1  &  5  & -2  \\
   2  &  4  & -4  \\
   2  & -2  &  2  
\end{array} \right)
\left( 
\begin{array}{r} 
c^{\text{zb}}_{11} \\
c^{\text{zb}}_{12} \\
c^{\text{zb}}_{44}
\end{array} \right ).
\label{eq:LinearMartin}
\end{equation}
It is already reasonable approximation, however, one can further improve the
transformation by adding so called internal strain (IS) nonlinear correction
to the Eq.~(\ref{eq:LinearMartin}). It affects only
three coefficients, namely, $c^{\text{w}}_{11}$, $c^{\text{w}}_{12}$,
$c^{\text{w}}_{44}$, and the corrected values can be expressed as
\begin{equation}
\left(
\begin{array}{r} 
  c^{\text{w}}_{11}  \\ 
  c^{\text{w}}_{33}  \\ 
  c^{\text{w}}_{12}  \\
  c^{\text{w}}_{13}  \\
  c^{\text{w}}_{44}  
\end{array} \right)
=
\left(
\begin{array}{r} 
  c^{\text{w0}}_{11}  \\ 
  c^{\text{w0}}_{33}  \\ 
  c^{\text{w0}}_{12}  \\
  c^{\text{w0}}_{13}  \\
  c^{\text{w0}}_{44}  
\end{array} \right)
+
\underbrace{
\left (
\begin{array}{c} 
  -\Delta^2/c^{\text{w0}}_{44}  \\ 
  0  \\ 
  \phantom{-}\Delta^2/c^{\text{w0}}_{44}  \\
  0  \\
  -\Delta^2/c^{\text{w0}}_{11}  
\end{array} \right),}_{\textrm{\bf IS}}
\label{eq:ISMartin}
\end{equation}
where $\Delta= 1/3 \sqrt{2} \, [1, -1, -2] [ \, c_{11}, c_{12}, c_{44}]^{T}$,
as derived by Martin. \cite{Martin1972,Martin1979}
To benchmark the presented approach for nitrides, we first cross-checked the
experimental values for wurtzite binaries AlN, GaN, and InN with those obtained
by the above transformation from our theoretical results for zinc blende
crystals. \cite{Lopuszynski2010} The comparison is presented in
Table~\ref{tab:CijBinary}.  Overall the agreement of our prediction with both
experimental findings (see Madelung tables \cite{Madelung2004}) as well as
recommended values combined from theory and experiment by Vurgaftman and Meyer
\cite{Vurgaftman2003} is good. One can note that already predictions based
solely on Eq. (\ref{eq:LinearMartin}) are very reasonable, even when the
nonlinear IS term given by Eq. (\ref{eq:ISMartin}) is neglected. The IS
correction is the most important for $c_{12}$ constant in examined materials.

\begin{table}[!ht]
\caption{Comparison of $c_{ij}$ for AlN, GaN, and InN obtained in this
work with experimental values provided in Madelung tables \cite{Madelung2004}
and with recommended values provided in the review paper of Vurgaftman and 
Meyer. \cite{Vurgaftman2003} \label{tab:CijBinary}}
\begin{ruledtabular}
\begin{tabular}{lrrrrrl}
      & $c_{11}$ & $c_{12}$ & $c_{13}$ & $c_{33}$ & $c_{44}$ & \\
\hline
  \dummy {\bf AlN} 
  & 411 & 149 &  99 & 389 & 125 & experimental values \cite{Madelung2004} \\ 
  & 396 & 137 & 108 & 373 & 116 & recommended values  \cite{Vurgaftman2003} \\
  & 373 & 119 & 101 & 391 & 108 & this work,  without IS \\
  & 366 & 126 &     &     & 107 & this work,  with IS \\
\hline
  \dummy {\bf GaN}
  & 377 & 160 & 114 & 209 &  81 & experimental values  \cite{Madelung2004} \\ 
  & 390 & 145 & 106 & 398 & 105 & recommended values \cite{Vurgaftman2003} \\ 
  & 337 & 113 &  97 & 353 &  95 & this work,  without IS \\
  & 331 & 119 &     &     &  94 & this work,  with IS \\
\hline
  \dummy {\bf InN} 
  & 190 & 104 & 121 & 182 &  10 & experimental values  \cite{Madelung2004} \\ 
  & 223 & 115 &  92 & 224 &  48 & recommended values \cite{Vurgaftman2003} \\
  & 211 &  95 &  86 & 220 &  48 & this work,  without IS \\
  & 207 &  99 &     &     &  47 & this work,  with IS \\
\end{tabular}
\end{ruledtabular}
\end{table}

\section{Elastic Constants in Wurtzite Alloys \label{sec:Results}} 
\begin{figure}[H]
\begin{minipage}{\textwidth}
\centering
\includegraphics[width=0.42\textwidth]{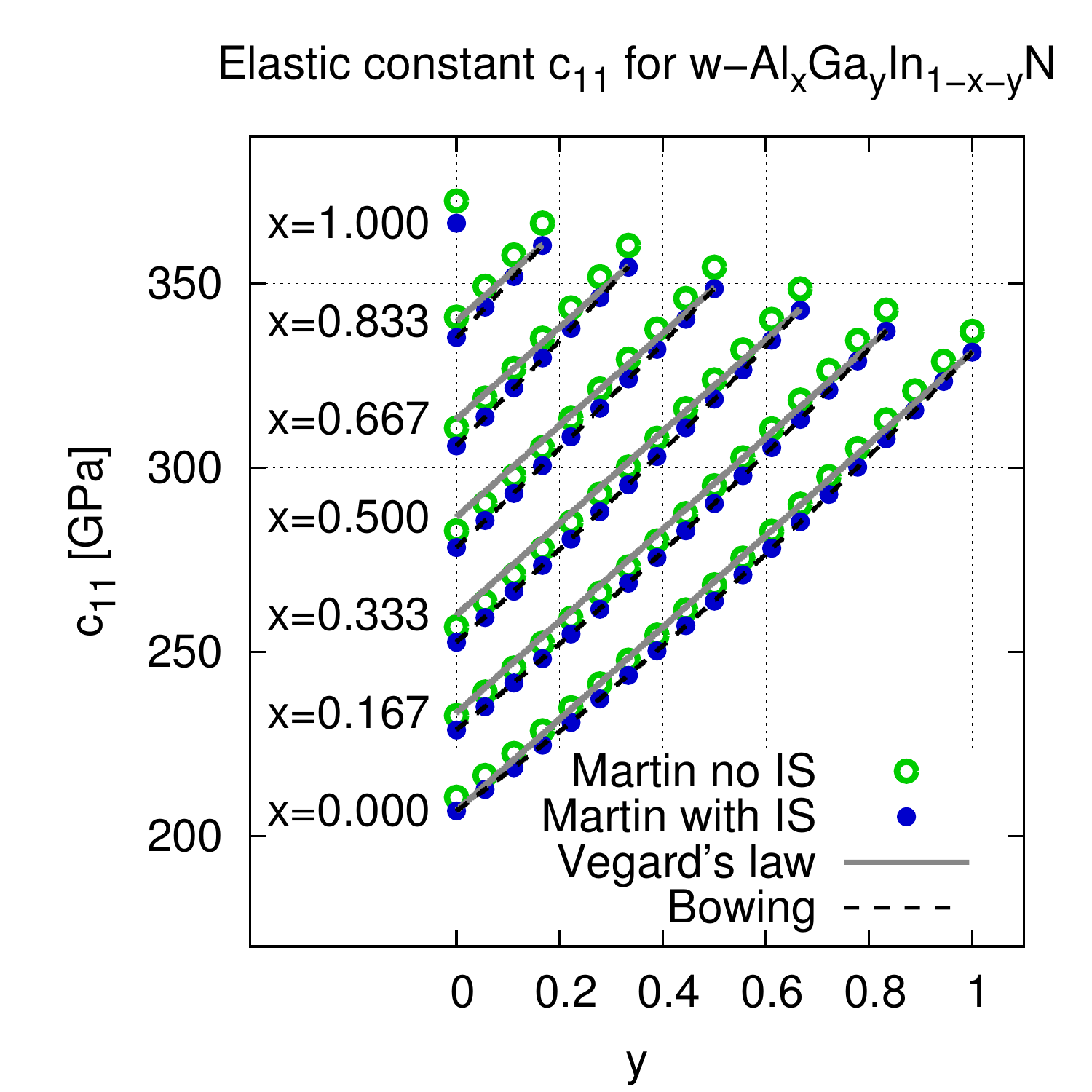}
\includegraphics[width=0.42\textwidth]{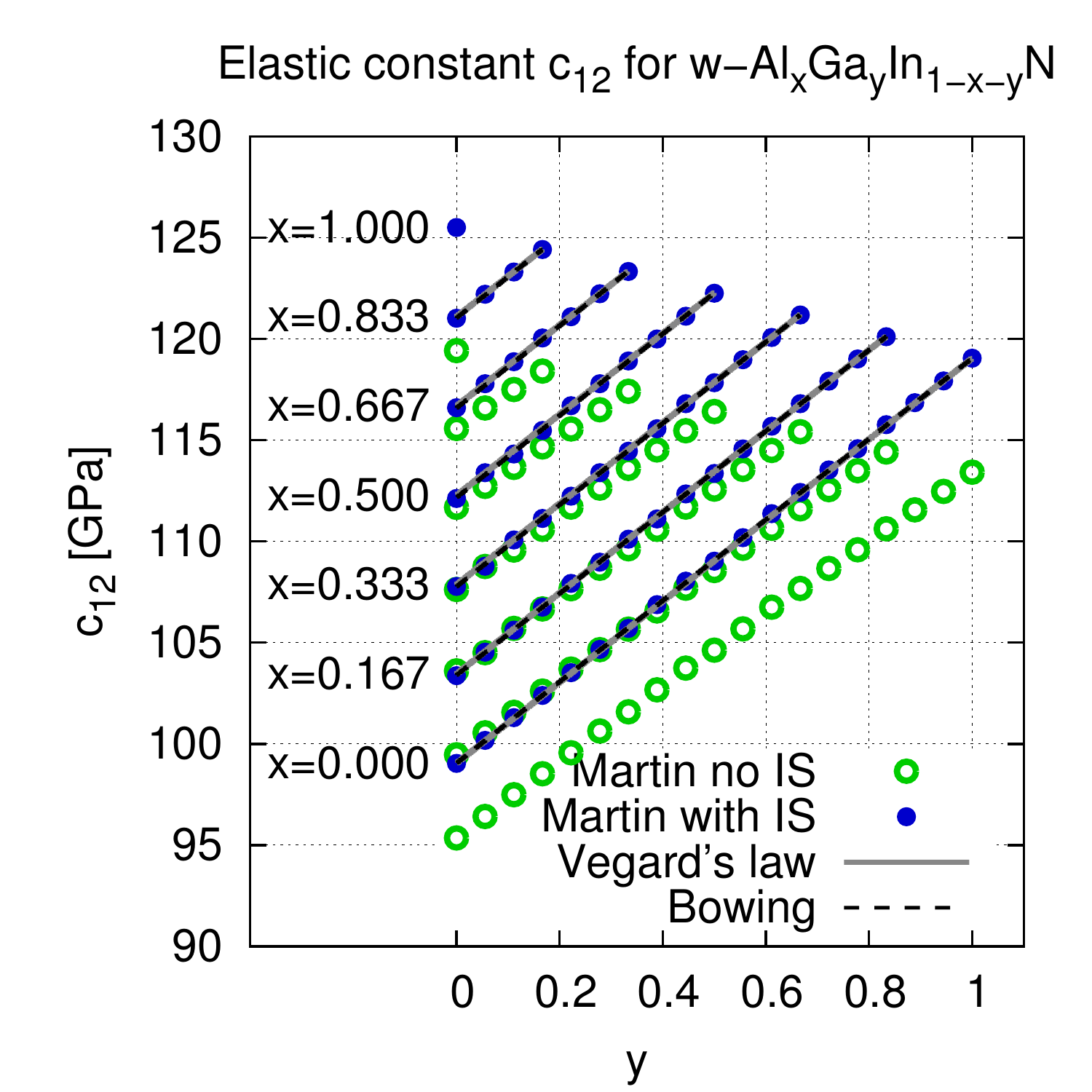}
\includegraphics[width=0.42\textwidth]{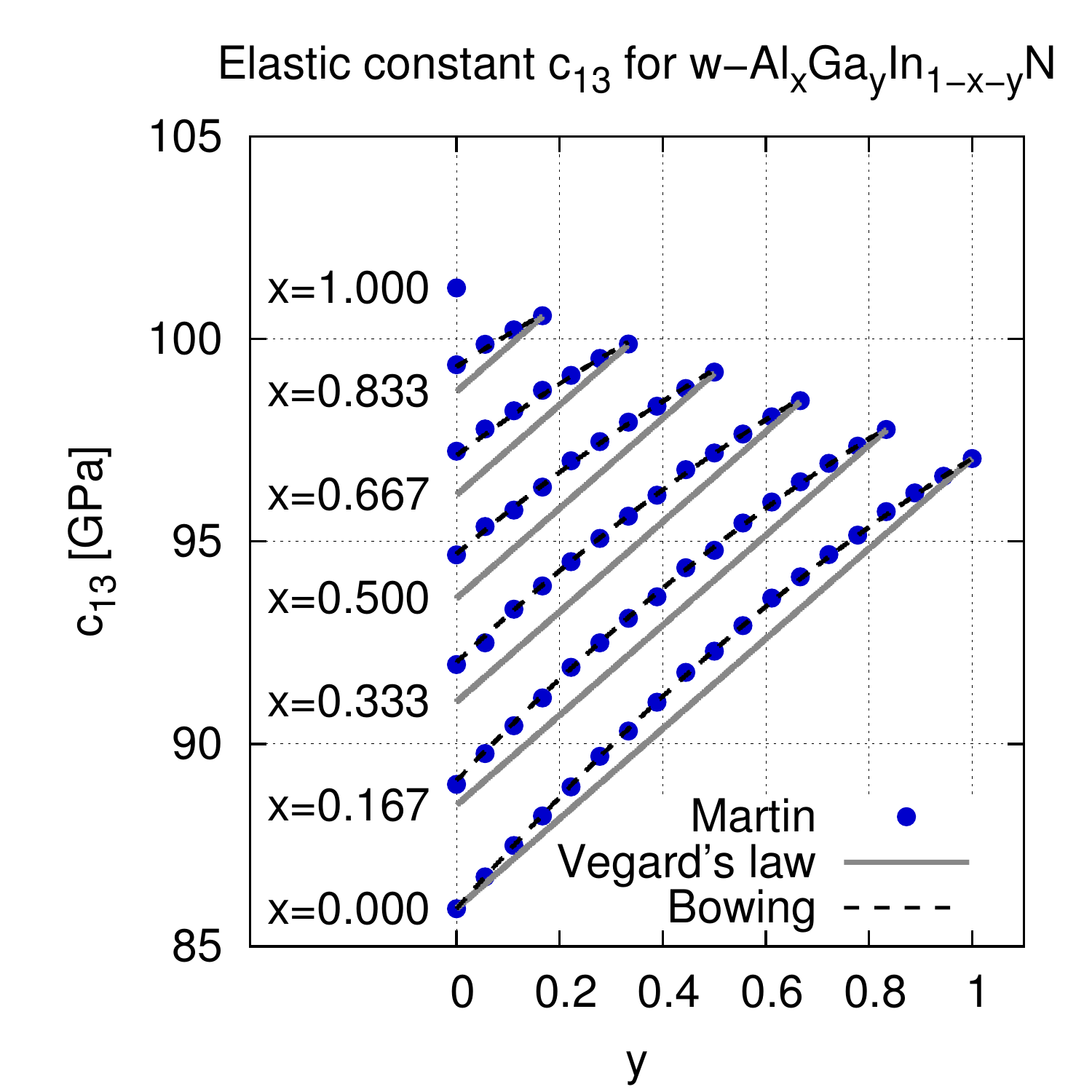}
\includegraphics[width=0.42\textwidth]{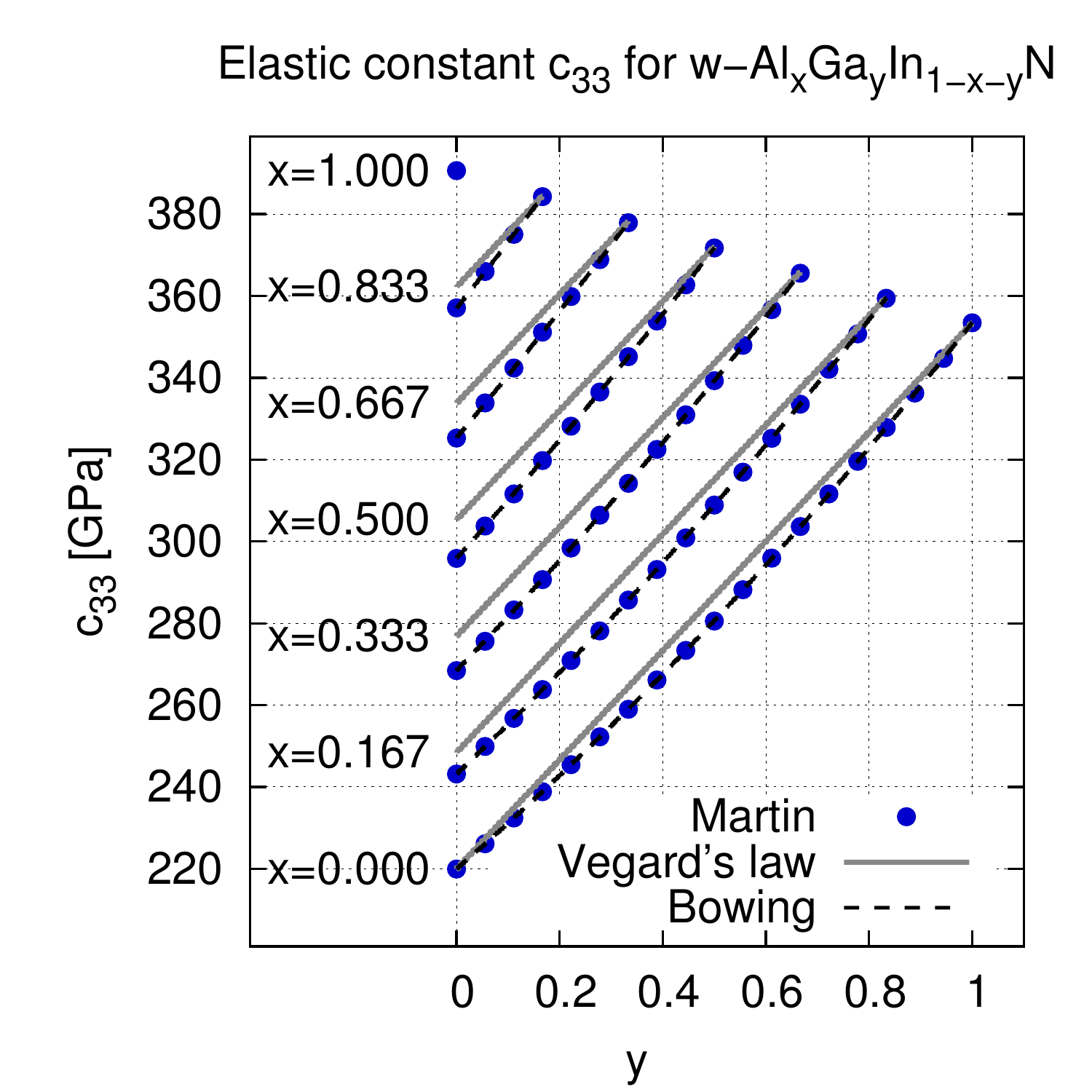}
\includegraphics[width=0.42\textwidth]{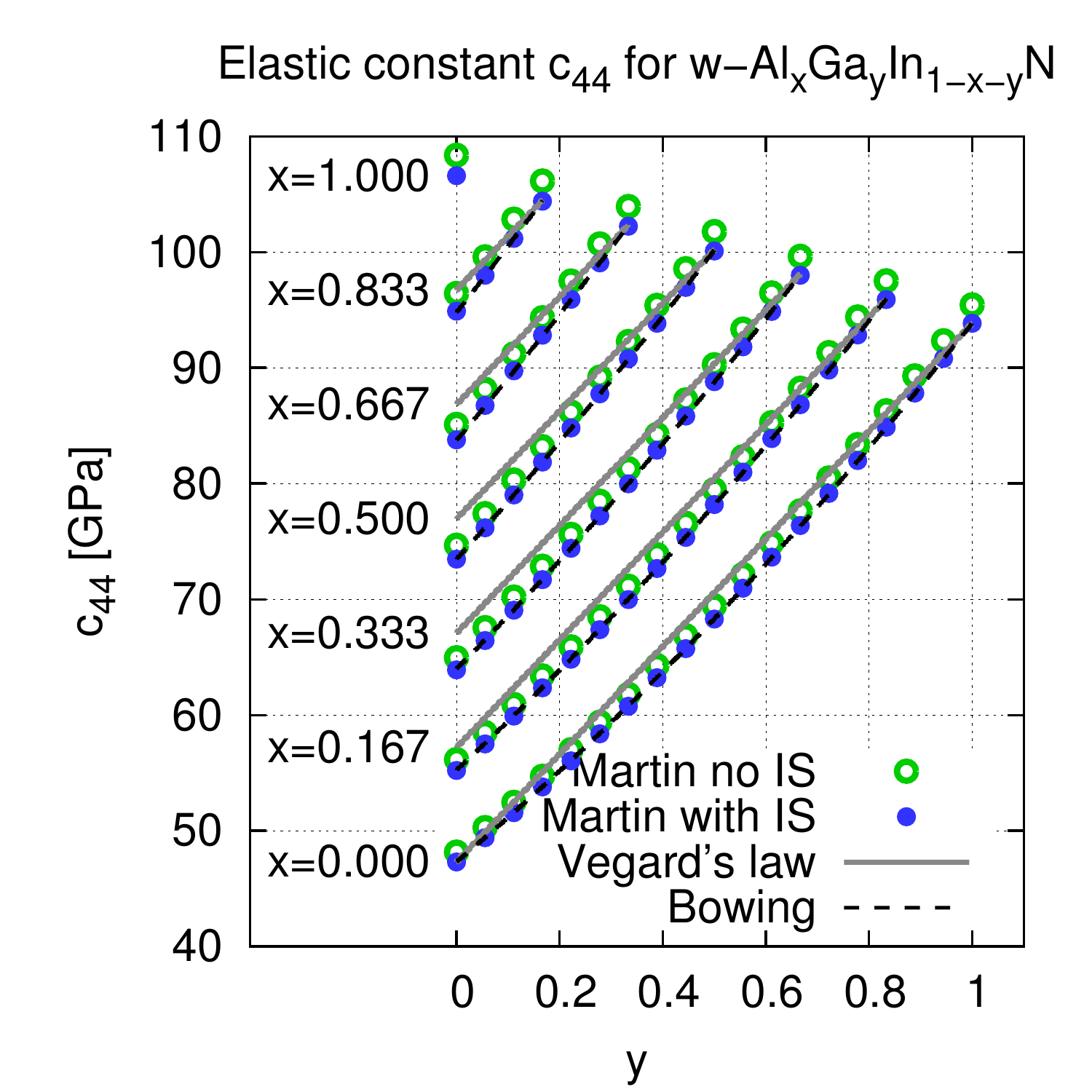}
\caption{Dependence of elastic constants $c_{ij}$ on composition 
for random wurtzite \AlGaInN alloys. For $c_{11}$, $c_{12}$, and 
$c_{44}$, both the results with internal strain contribution (IS)
and without it are presented. }
\label{fig:CijAlloys}
\end{minipage}
\end{figure}

After testing the approach on binary materials, we now focus on the case of
\AlGaInN alloys.  We apply the Martin transformation to the zinc blende values
resulting from our VFF calculations. \cite{Lopuszynski2010} The elastic
constants $c_{ij}$ as functions of AlN and GaN concentrations, $x$ and $y$
respectively, in quaternary \AlGaInN are displayed in
Fig.~\ref{fig:CijAlloys}. It turns out that in almost all cases the deviations
from linear Vegard-like law are present, similarly to the zinc blende phase. The
magnitude of the effect is not very large, reaching its maximum of $4.7 \%$ for
the $c_{44}$ case.  To describe accurately the observed dependencies, one has to
add the quadratic bowing contribution $\Delta c_{ij}(x,y)$  to Vegard-like law. 
The exact functional forms of this term for every considered elastic modulus is
depicted in Table~\ref{tab:CijAlloys}. With this correction the data are
reproduced with the accuracy of about $0.2\%$. One can also notice that
$c_{11}(x,y)$, $c_{33}(x,y)$, and $c_{44}(x,y)$ are sublinear. The $c_{12}(x,y)$
is well described by the linear model (Vegard-like law), the maximum observed
deviation is $0.2\%$. This is similar to the case of zinc blende nitride alloys,
where all dependencies were sublinear and the $c_{12}$ constant also exhibited
the smallest deviation from linearity. In contrast to that, the elastic constant
$c_{13}$ exhibits superlinear trend. Such a behavior was not observed for any
constant in the zinc blende case.  Another interesting issue here is the role of
nonlinear IS correction.  It turns out that for \AlGaInN alloys it has the
largest influence on $c_{12}(x,y)$ constant. For other constants affected by
IS, namely $c_{11}(x,y)$ and $c_{44}(x,y)$, the contribution is lower. In
addition, our simulations reveal that IS correction generally decreases
slightly the coefficients of bowing function.
As a final remark concerning our results for general quaternary \AlGaInN, let us
underline that the data for zinc blende nitrides taken as an input to this
calculations assume uniform random distribution of cations in the cationic
lattice. Therefore, if some kind of clustering occurs in the samples, this
effect is not taken into consideration.

Our results could be also compared with recent theoretical findings of
{\L}epkowski and Gorczyca for ternary \GaInN and \AlInN. \cite{Lepkowski2011}
They carried out their calculations for small wurtzite supercells containing 32
atoms, however, they used very accurate interaction description on the level of
quantum mechanical density functional theory (DFT). Interestingly, in order to
estimate the influence of clustering, for each composition they computed two
configurations --- uniform (with In atoms spread evenly throughout the cell) and
clustered (where In atoms where arranged close together). For ternary
nitride alloys $\text{A}_x \text{B}_{1-x}\text{N}$, the bowing term $\Delta
c_{ij}$ takes on very simple form
\begin{equation}
\Delta c_{ij}(\text{A}_x \text{B}_{1-x}\text{N}) = b\;x (1-x)
\end{equation}
having only one parameter $b$. Therefore, it is natural to use this bowing
parameter as a figure of merit for comparison of the two approaches.

\clearpage
\onecolumngrid
\twocolumngrid
\clearpage

The $b$ values obtained both from our calculations (extracted from
Table~\ref{tab:CijAlloys}) and those provided by {\L}epkowski and Gorczyca
\cite{Lepkowski2011} are compared in Table~\ref{tab:TernaryBowingComparison}.
Both clustered and uniform cases from Ref.~[\onlinecite{Lepkowski2011}] are
included for completeness.  Our results, however,  correspond to the uniform
case, since the underlying calculations for zb-\AlGaInN were obtained for even
distribution of cations in the sample.
\begin{table}[!t]
\caption{Results of fits to the dependence of elastic constants
         on composition for wurtzite \AlGaInN alloys. \label{tab:CijAlloys}}
\begin{ruledtabular}
{\small
\begin{tabular}{llcc}
\rule{0pt}{3ex}
         & Result of fit for $c_{ij}$
         & \multicolumn{2}{c}{Difference} \\
         &
         & \multicolumn{2}{c}{[GPa]} \\
\hline
$c_{11}$ & \dummy Vegard's law      &        &       \\
         & $206.81 +159.61\,x +124.63\,y$
         & 8.2 & (3\%) \\
         & With bowing term $ \Delta c_{11}$
                 &        &       \\
\dummyBtm         
         & $-32.90\,x(1-x) +52.97\,xy -20.89\,y(1-y)$
         & 0.2 & (0.1\%) \\
\hline
$c_{12}$ & \dummy Vegard's law      &        &       \\
\dummyBtm         
         & $99.04 +26.46\,x +20.00\,y$
         & 0.2 & (0.2\%) \\
\hline
$c_{13}$ & \dummy Vegard's law      &        &       \\
         & $85.93 +15.33\,x +11.12\,y$
         & 1.2 & (1.2\%) \\
         & \dummy With bowing term $ \Delta c_{13}$
                 &        &       \\
\dummyBtm 
         & $4.39\,x(1-x) -7.36\,xy +3.31\,y(1-y)$
         & 0.2 & (0.2\%) \\
\hline
$c_{33}$ & \dummy Vegard's law      &        &       \\
         & $219.93 +170.74\,x +133.52\,y$
         & 9.5 & (3.2\%) \\
         & \dummy With bowing term $ \Delta c_{33}$
                 &        &       \\
\dummyBtm
         & $-37.73\,x(1-x) +60.62\,xy -24.06\,y(1-y)$
         & 0.2 & (0.1\%) \\
\hline
$c_{44}$ & \dummy Vegard's law      &        &       \\
         & $47.31 +59.31\,x +46.55\,y$
         & 3.5 & (4.7\%) \\
         & \dummy With bowing term $ \Delta c_{44}$
                 &        &       \\
         & $-13.98\,x(1-x) +22.79\,xy -9.06\,y(1-y)$
         & 0.1 & (0.1\%) \\
\end{tabular} 
}
\end{ruledtabular}
\end{table}
It is interesting to notice that, even though the compared results were obtained
using very different methods, the qualitative agreement between them is very
good. Both approaches predict that significant sublinear behavior can be
expected for $c_{11}$, $c_{33}$, and $c_{44}$. The $c_{13}$ is expected to
exhibit slight superlinear tendency, whereas $c_{12}$ is the closest to
linearity in both models. Generally, the bowing coefficients obtained from
\abinitio calculations \cite{Lepkowski2011} are larger than in our VFF model.
The change to clustered distribution in the DFT modeling further amplifies the
difference. When it comes to the sources of this quantitative disagreement, both
approaches carry certain methodological shortcomings --- our approach is based
on a simple force field, but includes data from large supercell containing over
46 thousand atoms, which diminishes the finite size effects. The approach of
{\L}epkowski and Gorczyca \cite{Lepkowski2011} is based on the density
functional theory ansatz, so the interactions in that case are described very
accurately. However, employed small supercell carries systematic artificial
periodicity and implies certain kind of ordering. Interestingly, for the results
gathered in Table \ref{tab:TernaryBowingComparison}, one observes a clear trend
--- the higher the degree of ordering the larger the magnitude of bowing 
parameter $b$. Our results contain the smallest degree of ordering 
(large random supercell), so the observed $b$ values are the lowest.

\begin{table}
\newcommand{\tspA}{\hspace{1.5em} }
\newcommand{\tspB}{\hspace{1.5em} }
\newcommand{\tspC}{\hspace{1.5em} }
\caption{Comparison of results presented in this work with \abinitio
         calculation for ternary \GaInN and \AlInN obtained in 
         [\onlinecite{Lepkowski2011}]. \label{tab:TernaryBowingComparison}}
\begin{ruledtabular}
\begin{tabular}{crrr}
  Bowing 
  & \multicolumn{1}{c}{This work}    
  & \multicolumn{2}{c}{Previous work \cite{Lepkowski2011}} 
\\
  & \multicolumn{1}{c}{uniform}      
  & \multicolumn{1}{c}{uniform}  
  & \multicolumn{1}{c}{clustered} \\
\hline
\dummySml \GaInN      &           &           &            \\
\dummySml
$b(c_{11})$ &  -21\tspA &  -60\tspB &  -100\tspC \\ 
$b(c_{12})$ &    0\tspA &  -14\tspB &   -43\tspC \\
$b(c_{13})$ &    3\tspA &    4\tspB &     5\tspC \\
$b(c_{33})$ &  -24\tspA &  -71\tspB &     1\tspC \\
\dummyBtm
$b(c_{44})$ &   -9\tspA &  -16\tspB &   -35\tspC \\
\hline
\dummySml \AlInN      &           &           &            \\
$b(c_{11})$ &  -33\tspA &  -80\tspB &  -141\tspC \\
$b(c_{12})$ &    0\tspA &   -9\tspB &   -47\tspC \\
$b(c_{13})$ &    4\tspA &    3\tspB &    11\tspC \\
$b(c_{33})$ &  -38\tspA &  -25\tspB &    93\tspC \\
$b(c_{44})$ &  -14\tspA &  -35\tspB &   -70\tspC \\
\end{tabular}
\end{ruledtabular}
\end{table}

\section{Summary \label{sec:Summary}}
In this paper, the dependence of elastic constants on composition has been
studied theoretically for wurtzite random quaternary \AlGaInN.  It turns out
that all $c_{ij}(x,y)$, except for $c_{12}(x,y)$, deviate from the linear
dependence on composition, which is commonly employed to estimate
elastic properties of alloys. This deviation, however, is not very large,
usually around a few percent. We provide accurate quadratic fits to obtained
dependencies $c_{ij}(x,y)$ including this bowing effect.  Our calculations
reveal that for $c_{11}$, $c_{33}$, and $c_{44}$ linear model overestimates the
calculated moduli. On the other hand, in the case of $c_{13}$ the Vegard-like
law leads to underestimation of the material stiffness. The obtained results
agree qualitatively with recently published DFT results for elastic constants in
ternaries \GaInN and \AlInN. \cite{Lepkowski2011} Even though the described
nonlinearities in composition dependence of $c_{ij}$ are not very large, we
believe that the awareness of this effect could be useful in modeling nitride
heterostructures using continuous or coarse-grained models.

\section*{Acknowledgements}
M\L{} acknowledges the support of the European Union within European Regional
Development Fund, through the grant Innovative Economy
(POIG.01.01.02-00-008/08). JAM acknowledges the support of the SiCMAT
Project financed under the European Founds for Regional Development
(Contract No. UDA-POIG.01.03.01-14-155/09).

\bibliographystyle{apsrev4-1}
\bibliography{biblio}
\end{document}